\documentclass[twocolumn, preprintnumbers,amsmath,amssymb]{revtex4}
\usepackage{epsfig}

\newcommand{\beq}{\begin{equation}}
\newcommand{\eeq}{\end{equation}}
\newcommand{\bes}{\begin{subequations}}
\newcommand{\ees}{\end{subequations}}
\newcommand{\bea}{\begin{eqnarray}}
\newcommand{\eea}{\end{eqnarray}}
\newcommand{\ba}{\begin{array}}
\newcommand{\ea}{\end{array}}
\newcommand{\beqn}{\begin{eqnarray*}}
\newcommand{\eeqn}{\end{eqnarray*}}

\newcommand{\f}[2]{\frac{#1}{#2}}
\newcommand{\g}{\gamma}
\newcommand{\ep}{\epsilon}

\newcommand{\Se}{\Sigma}
\newcommand{\om}{\omega}

\newcommand{\la}{\langle}
\newcommand{\ra}{\rangle}
\newcommand{\mT}{\mathcal{T}}

\def\nn{\nonumber}

\begin{document}

\title{Thermoelectric phenomena in disordered open quantum systems}
\author{Dibyendu Roy and Massimiliano Di Ventra}
\affiliation{Department of Physics, University of California-San Diego, La Jolla, CA 92093}
\date{\today}

\begin{abstract}
Using a stochastic quantum approach, we study thermoelectric transport phenomena at low temperatures in disordered electrical systems connected to external baths. We discuss three different models of one-dimensional disordered electrons, namely the Anderson model of random on-site energies, the random-dimer model and the random-hopping model - also relevant for random-spin models. We find that although the asymptotic behavior of transport in open systems is closely related to that in closed systems for these noninteracting models, the magnitude of thermoelectric transport strongly depends on the boundary conditions and the baths spectral properties. This shows the importance of employing theories of open quantum systems in the study of energy transport.
\end{abstract}

\maketitle
Thermoelectric phenomena are attracting considerable attention both theoretically and experimentally, due to their fundamental unsolved aspects as well as
their impact in energy conversion technology~\cite{Bell08Dubi}. Here, we study thermoelectricity in disordered systems and the role of localization~\cite{Anderson58}, focusing mainly on energy transport by electrons. In real solids, electrical transport at finite temperatures is also controlled by inelastic electron-phonon scattering. Thus, we will only consider energy transport by electrons at low temperatures in the quantum regime. Quite often, despite their intrinsic open quantum system character, thermoelectric phenomena in disordered electrical systems are described within the framework of closed quantum systems, namely systems that do not exchange particles and/or energy with the environment~\cite{Chester61, Castellani87}. In fact it has been shown that energy transport in open disordered harmonic chains  is dependent not just on the properties of the system, but also on the baths connected to it~\cite{Dhar01}, as well as the boundary conditions~\cite{RoyDhar08}. Also there is a large class of driven disordered lattice  gas models of particles hopping on a lattice and interacting through hard-core exclusion where the nonequilibrium transport properties depend on the boundary conditions~\cite{Evans97, Barma06}. It is then natural to ask whether there are other fundamental properties of thermoelectric phenomena
that are not captured by closed-system theories and an open quantum system approach is necessary.

 To the best of our knowledge there are only a few microscopic studies (without assuming an explicit functional form of electrical conductivity) on thermoelectric transport properties in disordered closed systems \cite{VanLangen98, Romer03}. Recently, energy transport by electrons in disordered chains has been studied within an open quantum system approach~\cite{Dubi09}, but this study does not address the thermodynamic limit of system size and thus cannot be directly compared with the results of closed system theories. We therefore study thermoelectric  transport in a few models of open one-dimensional noninteracting disordered systems. Our main result is that although the asymptotic nature of transport at low temperatures in open systems is closely related to that in closed systems, the magnitude of thermoelectric transport depends on the boundary conditions and the baths' properties.

We employ a stochastic (Langevin) approach to investigate steady-state charge and energy transport in disordered noninteracting tight-binding lattices. This formalism has been extensively applied to study thermal transport in classical disordered lattices \cite{Rubin71, Dhar01} as well as quantum electrical \cite{DharSenRoy06} and phononic \cite{DharRoy06} systems. Its
main advantage is that one can explicitly consider the effect of baths and system-bath coupling. In this paper, we consider only non-interacting electrons. Interactions could be included
by, e.g., working with a stochastic Schr\"odinger equation in the context of time-dependent current-density functional theory, as it has been recently suggested in Ref.~\cite{Dagosta07-08}.
However, we leave this study for future investigations.

We consider three different one-dimensional (1D) tight-binding models of disordered electrons, namely, the Anderson model of random on-site energy (diagonal disorder) \cite{Anderson58}, the random dimer model (short-range correlated disorder) \cite{Dunlap90}, and the random hopping model (off-diagonal disorder) \cite{Theodorou76}. All states in the 1D independent random on-site energy Anderson model (AM) are exponentially localized for any strength of disorder \cite{Mott61}; thus there is no localization-delocalization transition in 1D. Now, one can have a localization-delocalization transition in 1D by introducing correlations (short or long range) in the random variables. The random dimer model (RDM) is the simplest example of that, where one or both of the two possible random on-site energies $\epsilon_a$ and $\epsilon_b$ are random in pairs. It has been shown that when both site energies appear in pairs there exist two real critical points with critical energies $\epsilon_a$ and $\epsilon_b$ if $|\epsilon_a-\epsilon_b| \le 2t$, where $t$ is the constant hopping strength between sites \cite{Sedrakyan04}. The random hopping model (RHM) is an example of a 1D model where a delocalized state appears at the band center even without any correlation in the randomness \cite{Theodorou76}. The last model has many common features with a wide class of random spin chains such as random $XY$ spin chains. All of the above results for the different disordered models have been derived for closed systems. Here we are particularly interested to know how these results are affected by the coupling with baths and what consequences we should expect on thermoelectric phenomena.

The full system consists of a disordered wire of $N$ sites and two infinite baths being connected to the wire at the two ends. The Hamiltonian is
\begin{equation}
\mathcal{H} = \mathcal{H}_W + \mathcal{H}_R^L + \mathcal{H}_R^R + \mathcal{V}_{WR}^L + \mathcal{V}_{WR}^R ~\label{ham}
\end{equation}
where $\mathcal{H}_W = -\sum_{l=1}^{N-1}t_l(c^{\dag}_l c_{l+1}+
 c^{\dag}_{l+1} c_l  ) +\sum_{l=1}^{N}\ep_l~c^{\dag}_l c_{l}$, $
\mathcal{H}_R^i = -\gamma_i\sum_{\alpha=1}^{\infty}~( c^{i\dag}_{\alpha}
 c^i_{\alpha+1} +c^{i\dag}_{\alpha +1} c^i_{\alpha})$ with $i=L,R$, and
$\mathcal{V}_{WR}^L = - \gamma'_L ( c^{L\dag}_{1}c_1 + c^{\dag}_1c^L_{1} ),\mathcal{V}_{WR}^R = - \gamma'_R( c^{R\dag}_{1} c_N + c^{\dag}_N c^R_{1} )$.

Here $ c_l$,~$c_\alpha^L $ and $ c_\alpha^R $ denote operators on the wire, the left and the right baths, respectively. The Hamiltonian of the wire is
 denoted by $\mathcal{H}_W$, that of the left (L) and the right (R) bath
 by $\mathcal{H}^i_R$ (with $i=L,R$); the tunneling Hamiltonian between the wire and the left (right) bath is $\mathcal{V}^L_{WR}~(\mathcal{V}^R_{WR})$.  The tunneling from the disordered wire to the baths is controlled by the parameters $\gamma'_i$. The AM of diagonal disorder is defined by constant electron hopping, i.e., $t_l=t$ for $l=1,2,..N-1$, and a distribution for the identically distributed independent random site energies $\epsilon_l$. In the RDM $t_l=t$ for $l=1,2,..N-1$ but there are only two values of on-site energy $\epsilon_l=\epsilon_a$ or $\epsilon_b$ which are assigned randomly in pairs with probabilities $p$ and $1-p$ respectively. Finally, in the case of the RHM, $\epsilon_l=0$ for $l=1,2,..N$, and the parameters $t_l$ are chosen from an independent random distribution.

We assume that each bath is in equilibrium at a specified temperature $T_i$ and chemical potential $\mu_i$ (with $i=L,R$) before coupling it with the wire. The coupling of the baths with the wire introduces noise and dissipation in the wire. We apply the stochastic approach following Ref.\cite{DharSenRoy06} to derive steady-state charge and energy current in the wire. Let us define $j^p$ and $j^u$ as the particle and the energy current density, respectively.
\bea
j^p&=&\f{1}{2 \pi}\int_{-\infty}^\infty d \om~ \mT_{1N}(\om)~(f_L-f_R) \label{jp}\\
j^u&=&\f{1}{2 \pi}\int_{-\infty}^\infty d \om ~\hbar \om~\mT_{1N}(\om)~(f_L-f_R) \label{ju}~,
\eea
where $\mT_{1N}=4 \pi^2 {\g'_L}^2 {\g'_R}^2 \rho_L(\om) \rho_R (\om) |G_{1N}|^2/\hbar^4$,
$\hat{G}=\hat{Z}^{-1}$, $\hat{Z}=\hat{\Phi}-\hat{\Se}(\om)$ and
\begin{equation}
\Phi_{lm}=(\omega-\f{\ep_l}{\hbar}) ~\delta_{l,m}+\f{t_l}{\hbar}\delta_{l,m-1}+\f{t_m}{\hbar}\delta_{l,m+1}.
\end{equation}

Here $\hat{\Phi}$ is the Hamiltonian matrix of the disordered wire and $\hat{G}$ is the full Green's function of the wire coupled with the baths. The self-energy correction $\hat{\Se}$, coming from the baths, is a $N \times N$ matrix whose only non-zero elements are $\Se_{11}$ and $\Se_{NN}$. In the following we set $\hbar=1$ and discuss the case in which  $\g_L=\g_R=\g$, $\g'_L=\g'_R=\g'$, which corresponds to symmetric coupling to the baths. In this case, $\Se^+_R(\om)=\Se^+_L(\om)=\Se(\om)={\g'}^2g_{1,1}(\om)$ where $g_{1,1}(\om)$ is the single particle Green's function of the isolated bath at the first site $\alpha=1$:
\bea
 g_{1,1}(\om)&=&\f{1}{\g}\big[ \f{ \om}{2\g}-i\big(1-\f{\om^2}{4{\g}^2}\big)^{1/2}\big] \nn
\eea
Also in Eqs.(\ref{jp},\ref{ju}), $f_i=1/\{exp[( \om-\mu_i)/k_B T_i]+1\}$ is the Fermi function of the $i$th bath and $\rho_i(\om)=-{\rm Im} [g_{1,1}(\om)]/\pi$ is the local density of states at the first site $(\alpha=1)$ on the $i$th bath. It can be shown that $\mT_{1N}(\omega)$ is the transmission coefficient of an electron from the left to the right bath through the disordered chain at energy $ \omega$. Interestingly for $\g=1$ (ideal baths) and $\g'=1$ (ideal contacts) the above formulation merges with the Landauer theory of transport.

We are interested in the system properties in the thermodynamic limit. We thus extend a technique originally developed by Dhar \cite{Dhar01} to determine steady-state thermal currents in a disordered harmonic chain connected to baths to the present case of disordered electrical open quantum systems. This technique is a generalization of the popular recursive Green's function method \cite{Romer03} to open systems. We are interested in finding the asymptotic system size $(N)$ dependence of the steady state $\la j^p \ra$ and $\la j^u \ra$, where $\la\cdots\ra$ denotes average over disorder realizations. Following \cite{Dhar01, RoyDhar08} we now separate out the wire and the bath contributions in $G_{1N}$ and write them explicitly:
\bea
|G_{1N}|^2&=&|\Delta_N(\om)|^{-2}\prod_{l=1}^{N-1}t_l^2~~~{\rm with} \label{gr} \\
\Delta_N(\om)&=&D_{1,N}-\Se(\om)(D_{2,N}+D_{1,N-1})+\Se^{2}(\om)D_{2,N-1}\nn
\eea
where $\Delta_N(\om)$ is the determinant of $\hat{Z}$ and $D_{l,m}$ is the determinant of the sub-matrix of $\hat{\Phi}$, beginning with the $l$th row and column, and ending with the $m$th row and column. We can numerically compute the elements $D_{l,m}$ efficiently by taking the product of the $2 \times 2$ random matrices $\hat{T}_l$:
\bea
&& \hat{D}=\left( \begin{array}{cc}
 D_{1,N} & -D_{1,N-1} \\ D_{2,N} & -D_{2,N-1} \end{array} \right)=\hat{T}_1
\hat{T}_2....\hat{T}_N \label{trans} \\
{\rm{where}}~~ &&\hat{T}_l=\left( \begin{array}{cc}
\om-\epsilon_l & -t^2_l \\ 1 & 0 \end{array}
\right){\rm{for}}~~l=1,2,..N-1 \nn \\ {\rm{and}}~~ &&\hat{T}_N=\left( \begin{array}{cc}
\om-\epsilon_l & -1 \\ 1 & 0 \end{array}
\right) \nn
\eea
After a  little algebra we find
\bea
|\Delta_N(\om)|^2&=&D^2_{1,N}+\f{\g'^8}{\g^4}D^2_{2,N-1}+\f{\g'^4}{\g^2}(D_{2,N}+D_{1,N-1})^2\nn \\
&-&\f{\g'^2}{\g^2}\om (D_{1,N}+\f{\g'^4}{\g^2}D_{2,N-1})(D_{2,N}+D_{1,N-1})\nn \\&+&\f{\g'^4}{\g^2}\Big(\f{\om^2}{\g^2}-2\Big)D_{1,N}D_{2,N-1}~. \label{det}
\eea
We compute $\la\mT_{1N}(\om)  \ra$  numerically using Eqs.(\ref{gr}-\ref{det}). The above formulation can be used to find steady-state currents at finite bias, but in this paper we are interested in the linear response regime at low temperatures, i.e., $\Delta \mu<<\mu,~\Delta T<<T$ and $k_BT<<\mu$ with $\Delta \mu=\mu_L-\mu_R,~\Delta T=T_L-T_R,~\mu=(\mu_L+\mu_R)/2$ and $T=(T_L+T_R)/2$. In the linear response regime we write the heat current as $j^h=j^u-\mu j^p$. We then find
\bea
j^p&=&\f{-1}{2\pi}[\mT_{1N}(\mu)\Delta \mu + \f{\pi^2k_B^2T}{3} \mT'_{1N}(\mu)\Delta T]~,\nn \\
j^h&=&\f{-1}{2\pi}[\f{\pi^2k_B^2T^2}{3}\mT'_{1N}(\mu)\Delta \mu + \f{\pi^2k_B^2T}{3} \mT_{1N}(\mu)\Delta T]~.\nn
\eea

$Results -$
\begin{figure}
\begin{center}
\includegraphics[width=8.0cm]{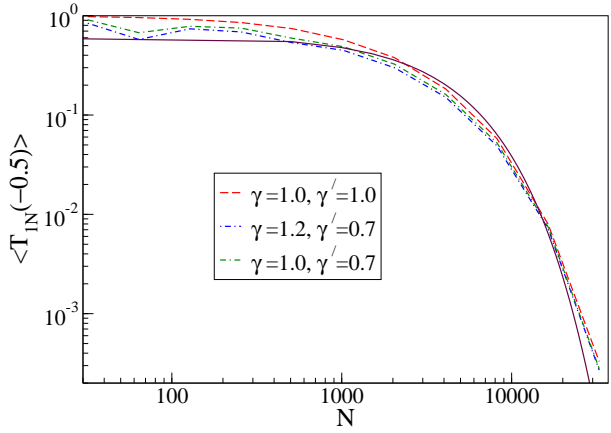}
\end{center}
\caption{(Color online) Plot of $\la\mT_{1N}(-0.5)\ra$ versus $N$ for three different sets of $\g$ and $\g'$ in the AM with $t=1$. $\epsilon_l$ at each site $l$ is drawn from a Gaussian distribution of mean zero and variance $\sigma^2=0.0025$. The full curve is asymptotic scaling of $\la \mT_{1N}(-0.5)\ra$ for all three sets of parameters.
}
\label{fig1}
\end{figure}
We thus see from the above results that particle and thermal conductances scale similarly even in the disordered open quantum systems and the Weidemann-Franz relation is valid in the
absence of interactions. We also note that the asymptotic size dependence of the electrical and thermal conductances depend on the asymptotic behavior of $\la\mT_{1N}(\mu)\ra$. Hereafter we will then focus on this quantity. We find that for independent random distribution of on-site disorder in the 1D AM, both particle current and heat current become exponentially small in the  asymptotic limit of system size in the closed as well as the open systems. Thus the behavior of $\la\mT_{1N}(\mu)\ra$ in the asymptotic regime is dominated by localization physics and there is clearly no diffusive behavior satisfying either Ohm's or Fourier's law. The asymptotic character of the eigenstates of disordered systems is quantified through the localization length $\xi(\om)$ which is defined by~\cite{DiVentrabook}
\bea
{\rm lim}_{N \to \infty}N^{-1}{\rm ln}\la \mT_{1N}(\om)\ra=-2/\xi(\om).\label{locl}
\eea
 For a disordered tight-binding chain of lattice constant $a$, with a random on-site potential $\epsilon_l$ at each site $l$ drawn from a Gaussian distribution of mean zero and variance $\sigma^2$, the localization length of the closed chain is given by $\xi(E_F)=2(a/\sigma^2)(4t^2-E_F^2)$, where $E_F$ is the chemical potential at zero temperature or the Fermi energy \cite{Dorokhov92}. Thus $\xi(E_F)= 3000$ for $\sigma^2=0.0025~t^2$ and $E_F=-0.5t$ with $a=1$. We compute numerically the localization length in the disordered closed and open chains for the same above parameters using the definition of Eq.(\ref{locl}). We find in numerics with $t=1$, $a=1$ (see Fig.\ref{fig1}) that $\xi(E_F)~(\simeq 6000)$ is the same for both the closed and the open chains ($\g'=0.7,~\g=1.2$). 
 We also find that the magnitude of $\la\mT_{1N}(\mu)\ra$ in the open chains falls rapidly from that of the closed chains for any changes of $\g$ and $\g'$ from the unity. This is expected and can be understood physically.  Any value of $\g$ and $\g'$ different from unity introduces extra scattering in the chain and thus reduces the magnitude of $\la\mT_{1N}(\mu)\ra$.

It has been shown in Ref.\cite{Sedrakyan04} that all states of the closed RDM are localized except at the two energies $\epsilon_a$ and $\epsilon_b$ which are real critical points with infinite $\xi$ if $|\epsilon_a-\epsilon_b| \le 2t$. In numerics with the closed systems we instead find that though $\la \mT_{1N}(\om)\ra$ at $\om=\epsilon_a ~{\rm or}~ \epsilon_b$ remains constant with increasing system size, it decays algebraically ($\la \mT_{1N}(\om)\ra \sim N^{-2}$) with system size for  $|\epsilon_a-\epsilon_b|\simeq 2t$. Thus it is hard to conclude convincingly whether the energies (i.e., $\om=\epsilon_a ~{\rm or}~ \epsilon_b$) are true critical points for $|\epsilon_a-\epsilon_b|\simeq 2t$. Interestingly,  the sample to sample fluctuations in $\la \mT_{1N}(\om)\ra$ for the latter case decay with increasing system sizes; while away from these energies (or in the AM) the sample to sample fluctuations in $\la \mT_{1N}(\om)\ra$ do not fall with increasing length. In the open systems (with $\g,\g' \ne 1$) we again find similar asymptotic behavior as the closed systems in numerics. Thus in open systems the transport is ballistic at $\om=\epsilon_a ~{\rm or}~ \epsilon_b$ for  $|\epsilon_a-\epsilon_b| < 2t$ and it shows power law dependence on length at $\om=\epsilon_a ~{\rm or}~ \epsilon_b$ for  $|\epsilon_a-\epsilon_b|\simeq 2t$.

\begin{figure}
\includegraphics[width=8.0cm]{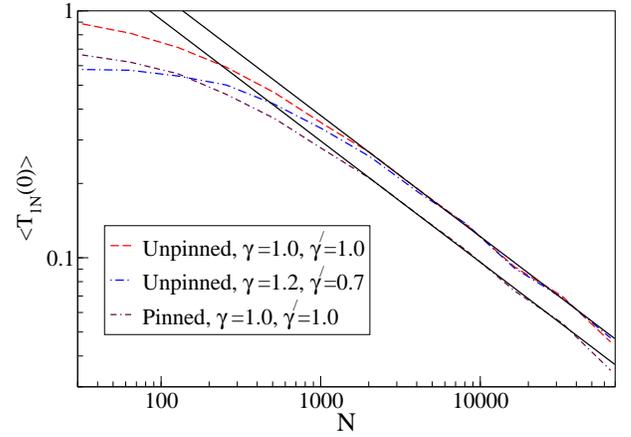}
\caption{(Color online) Plot of $\la \mT_{1N}(0)\ra$ versus $N$ in the pinned and unpinned RHM for different $\g,~\g'$ with $t=0.8$ and $\Delta t=0.2$. The two straight lines correspond to the asymptotic of $\la \mT_{1N}(0) \ra$. The sample to sample fluctuations in $\mT_{1N}(0)$ are quite large (on the order of $\la \mT_{1N}(0)\ra$) and do not decay with increasing realizations or system size.
}
\label{fig2}
\end{figure}
It has been argued from simple considerations that for the closed RHM, a state at the band center, $\om=0$, is extended and all other states are exponentially localized \cite{Theodorou76}. Here, again we find in numerics that $\la \mT_{1N}(0)\ra$ does not remain constant as a function of $N$ in the closed and open RHM, but it decays with increasing $N$. The asymptotic length dependence of $\la \mT_{1N}(0)\ra$ is given by $N^{-0.5}$ (see Fig.\ref{fig2}). We use a uniform distribution between $t$ and $t+\Delta t$ for $t_l$.  The RHM model is equivalent to a disordered linear chain of harmonic oscillators for closed systems \cite{Theodorou76, Dyson53}. Here, we wish to compare the open system results of the RHM with that of the random spring harmonic chains. Recently the authors of \cite{Gaul07} have investigated heat conduction in random spring quantum harmonic chains for shorter lengths; but they were not able to conclude about the length dependence of the disorder averaged steady-state thermal current $(\la J \ra)$ in the spring model. It has been already argued in Ref.~\cite{RoyDhar08} that the asymptotic length dependence of the classical and quantum thermal currents are similar in disordered harmonic chains.
Also the asymptotic length dependence of  $\la J \ra$ in the random spring harmonic chains, for two different models of baths (Rubin's baths: $\Sigma(\omega)= k\{1-m\omega^2/2k-i\omega (m/k)^{1/2}\left [1-m\omega^2/(4k)\right ]^{1/2}\}$ with $k$ the spring constant and $m$ the mass of the lattice site; and white-noise baths: $\Sigma(\omega)=-i\gamma\omega$) are similar to that of the random mass harmonic chains: $\la J \ra \sim N^{-0.5}$ for Rubin's baths and $\la J \ra \sim N^{-1.5}$ for white noise baths.  Interestingly, we see that the asymptotic length dependence of thermal currents in the Rubin's bath case is similar to that of the open RHM. 
We further put two pinning potentials $(\epsilon_1=\epsilon_N=1)$ at the two ends of the RHM to examine the effect of different boundary conditions. We find that the scaling of $\la \mT_{1N}(0)\ra$ remains the same as the unpinned case (see Fig.\ref{fig2}). Two external quadratic pinning potentials at the two ends of the disordered harmonic chain with Rubin's baths change the asymptotic length dependence of $\la J \ra$ to $N^{-1.5}$, thus showing a difference in energy transport by the RHM and the random spring harmonic chain. This can be understood as follows.  While energy transport in tight-binding chains mostly occurs by electrons at the chemical potential, the full band of conducting modes in harmonic chains carries energy. The external pinning does not affect the band center in the RHM; but breaks the translational invariance in the harmonic chains and pinches off the band of conducting modes from the zero frequency side, thus reducing $\la J \ra$.

 In conclusion, we have shown that the asymptotic nature of thermoelectric transport in the noninteracting disordered open systems is quite similar to that in the closed systems. However, the magnitude of thermal and electrical conductances is smaller in the open systems compared to that in the closed systems. In earlier studies the effect of coupling with baths has been included through a phenomenological lifetime due to inelastic scattering from the baths. This is done by energy continuation into the complex plane. Here, we have explicitly included the baths in our microscopic analysis. Since the technique used in this paper can be extended to higher dimensions following Ref.~\cite{Chaudhury09}, it would be interesting to analyze our results in 2D and 3D. It is known that the Anderson localization-delocalization transition in the random 3D AM will be smoothed in the presence of inelastic scattering (due to baths), but it will be interesting to check how the corresponding transport properties will be affected by explicit coupling with the baths near the transition. Experiments in disordered systems are mostly carried out in open configurations. In fact, many real disordered systems such as doped polyaniline, random semiconductor superlattices  \cite{Bellani99} and random antiferromagnetic spin chains \cite{Boucher96} are considered to have similarity with the RDM and the RHM. Therefore, we expect our results to be useful in understanding experiments in these systems.

This work has been funded by the DOE grant DE-FG02-05ER46204 and UC Laboratories.


\end{document}